\documentclass[12pt,a4paper,twocolumn]{narms}

\usepackage{psfig}
\usepackage{timesmt}   % use times typeface
\usepackage{chicago}   % <---------------------------------- MOD
\usepackage[small]{caption}
\usepackage{epsfig}
\usepackage{amsmath}
\usepackage{amssymb}
\usepackage{wasysym} 

\makeatletter
\renewcommand{\section}{\@startsection%
{section}%
{1}%
{0mm}%
{- \baselineskip}%
{0.15\baselineskip}%
{\normalfont\normalsize}}%

\renewcommand{\subsection}{\@startsection
{subsection}%
{2}%
{0mm}%
{-\baselineskip}%
{0.15\baselineskip}%
{\normalfont\normalsize}}%
\makeatother

\begin{document}
%\maketitle

\title{On a three-dimensional lattice approach for modelling corrosion induced cracking and its influence on bond between reinforcement and concrete}
\author{\large {Peter Grassl and Trevor Davies}\\
{\em Department of Civil Engineering, University of Glasgow, Glasgow, United Kingdom}\\
}
\date{}% No date.
\pagestyle{empty}
\thispagestyle{empty}
\abstract{ABSTRACT:
The present work involves the discrete modelling of corrosion induced cracking and its influence on the bond between reinforcement and concrete.
A lattice approach is used to describe the mechanical interaction of a corroding reinforcement bar, the surrounding concrete and the interface between steel reinforcement and concrete.
The cross-section of the ribbed reinforcement bar is taken to be circular, assuming that the interaction of the ribs of the deformed reinforcement bar and the surrounding concrete is included in a cap-plasticity interface model.
The expansion of the corrosion product is represented by an eigenstrain in the lattice elements forming the interface.
The lattice modelling approach is applied to the analysis of corrosion induced cracking and its influence of the bond strength.
The model capabilities are assessed by comparing results of analyses with those from unconfined pull-out tests reported in the literature.
Future work will investigate the influence of the stiffness of interface elements and the effect of lateral confinement on corrosion induced cracking.

\vspace{5mm}
Keywords: lattice, concrete, cracking, reinforcement, corrosion
\vspace{-5mm}
}

%%%%%%%%%%%%%%%%%%%%%%%%%%%%%%%%%%%%%%%%%%%%%%%%%%%%%%%%%%%%%%%%%%%

%  To kick in the changes.
% Make title here for two column format.
\maketitle

%\paragraph{Keywords: damage, concrete, mesh bias, fracture, isotropic, anisotropic, crack tracking}

\frenchspacing   % no double spaces after colon
                 % added by <W.Hennings@fz-juelich.de>

% Write some ascii text files called intro.tex, concept.tex, etc.
% TeX and LaTeX will look for the .tex subscript by default.

%hint: for citations according to Balkema style, use "\shortshortcite{key}"
%      instead of "\shortcite{key}

%%%%%%%%%%%%%%%%%%%%%%%%%%%%%%%%%%%%%%%%%%%%%%%%%%%%%%%%%%%%%%%%%%%

\section{INTRODUCTION}

The present work involves the modelling of corrosion induced cracking of reinforced concrete by means of a three-dimensional lattice approach.
Corrosion of reinforcement involves the transformation of steel into expansive rust.
If the expansion of rust is restrained, it results in radial pressure in the confining material. 
For reinforced concrete, this radial pressure and accompanying transverse tensile stresses may cause cracking \shortcite{AndAloMol93}. 
This cracking is not desirable, since it reduces the anchorage capacity of the reinforcement \shortcite{LeeNogTom02}.
Most of the anchorage capacity of deformed reinforcement is provided by ribs on the surface of the bar, which resist the slip between concrete and reinforcement by transferring inclined radial forces into the concrete~\shortcite{Tep79}.
The capacity of the concrete to resist these forces can be significantly reduced by the corrosion induced cracking.
Consequently, there is a considerable interest in developing models which can predict the mechanism of corrosion induced cracking and its influence on the bond capacity of reinforced concrete.

Many of the present models include the effect of corrosion induced cracking on bond by reducing the bond strength of the interface between concrete and steel \shortcite{LeeNogTom02}.
In these models, the relationship between the amount of rust and the reduction of bond strength is determined empirically.
Thus, these models are of limited validity for the prediction of the influence of corrosion on bond.

\begin{figure}
\begin{center}
\epsfig{file=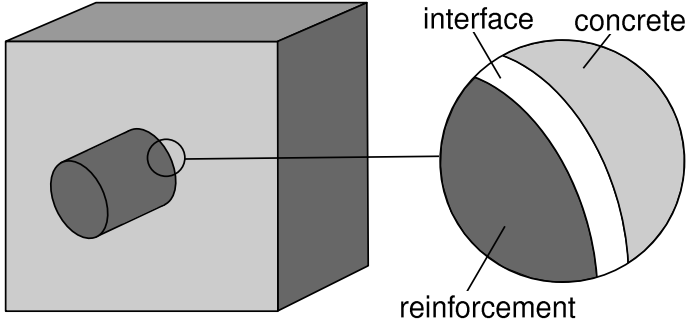,width=9cm}\\
\end{center}
\caption{Three-dimensional modelling of reinforced concrete: Concrete, reinforcement and bond between concrete and reinforcement are considered as individual phases. Left: Concrete cube (light grey) with reinforcement bar (dark grey); right: detail of interface between reinforcement and concerte.}
\label{fig:3DBond}
\end{figure}

Only very few models describe the expansion of the rust, the radial pressure and the transverse stresses on the concrete explicitly \shortcite{Lun05}.
These models have the potential to establish an analytical relationship between the expansion of rust, cracking and spalling.
They can be combined with realistic bond models \shortcite{Lun05a}, so that the influence of corrosion induced cracking on the bond capacity can be predicted without the need of empirical relationships.
However, this modelling framework is computationally demanding, since it requires three-dimensional modelling of the mechanical response of the concrete, the bond between reinforcement bar and concrete, and the reinforcement bar itself as shown in Figure~\ref{fig:3DBond}.

In earlier work, three-dimensional continuum based models were used to describe the cracking of concrete surrounding the reinforcement.
However, three-dimensional continuum modelling of cracking in concrete is challenging, since it is not straight-forward to include the localised deformations into the continuum description. Combined with the modelling of the bond between concrete and reinforcement, it can be exceedingly difficult.
This might explain the small number of models which are available based on this three-dimensional approach.
Discrete approaches, such as lattice and particle models, might be a favourable alternative to describe this discontinuous problem. Their potential to model corrosion induced cracking and its influence on bond is assessed in the present study.

Lattice approaches have been used successfully in the past to model the failure of concrete, as reported by \shortciteN{SchMie92b} and \shortciteN{BolSai98}.
The model by Bolander has been shown to accurately reproduce analytical solutions for elasticity and potential flow problems \shortcite{YipMohBol05,BolBer04}. Furthermore, it allows for the use of constitutive models, which are formulated by means of tractions and displacement jumps, as commonly used in interface approaches for concrete fracture \shortcite{CabLopCar06}. These have been shown to result in an element-size independent description of crack-openings. 
This lattice approach is used in the present study to describe the mechanical response of three phases, namely reinforcement, concrete and bond between reinforcement and concrete.
In this approach, the lattice elements do not represent the meso-structure of the materials~\cite{ZubBaz87}. Instead, they are used to discretise the continuum.
Thus, constitutive models are required for all three phases, which are described in more detail in the following sections.

\section{MODELLING APPROACH}
The present lattice approach for the modelling of corrosion induced cracking follows the framework developed by Bolander and his co-workers.
The nodes of the lattice are randomly located in the domain to be analysed, subject to the constraint of a minimum distance \shortcite{ZubBaz87}.
The arrangement of the lattice elements is determined from the edges of the tetrahedra of the Delaunay tessellation based on the randomly placed nodes.
The cross-sectional properties of these elements are obtained from the dual Voronoi tessellation of the same set of random nodes. 
For the interface between reinforcement and concrete, the lattice nodes are not placed randomly but at special locations,
such that the middle cross-sections of the lattice elements form the boundaries between the two phases \shortcite{BolBer04}.
The nodes of the lattice elements have six degrees of freedom, namely three translations and three rotations.
These degrees of freedom are related to three displacement and three rotation discontinuities at the centroid of the middle cross-section of the elements. 
The three rotation discontinuities are related to moments by elastic relationships.
The three displacement discontinuities are used in interface constitutive models to determine the corresponding tractions.
In the present study, an elastic interface model is used for the reinforcement. 
One limitation of the present lattice approach is that it only predicts Poisson's ratios of less than $1/4$ for 3D and less than $1/3$ for 2D.
This is restrictive for the 3D modelling of the steel reinforcement, which has a Poisson's ratio greater than $1/4$.
This limitation could be overcome by combining the lattice model with continuum tetrahedra as discussed for the 2D case by \shortciteN{GraBazCus06}. 
However, this approach is beyond the scope of the present study. 
The interface model for concrete is based on a combination of plasticity and damage, which describes the softening and reduction of the unloading stiffness in tension as well as the nonlinear hardening response in high confined compression \shortcite{Gra09a}. For the interface between concrete and reinforcement, a new cap-plasticity model is proposed which is described in more detail in the next section. 

\subsection{\em Elasto-plastic cap model for the bond between concrete and reinforcement}
\begin{figure}[h!]
\begin{center}
\begin{tabular}{c}
\epsfig{file=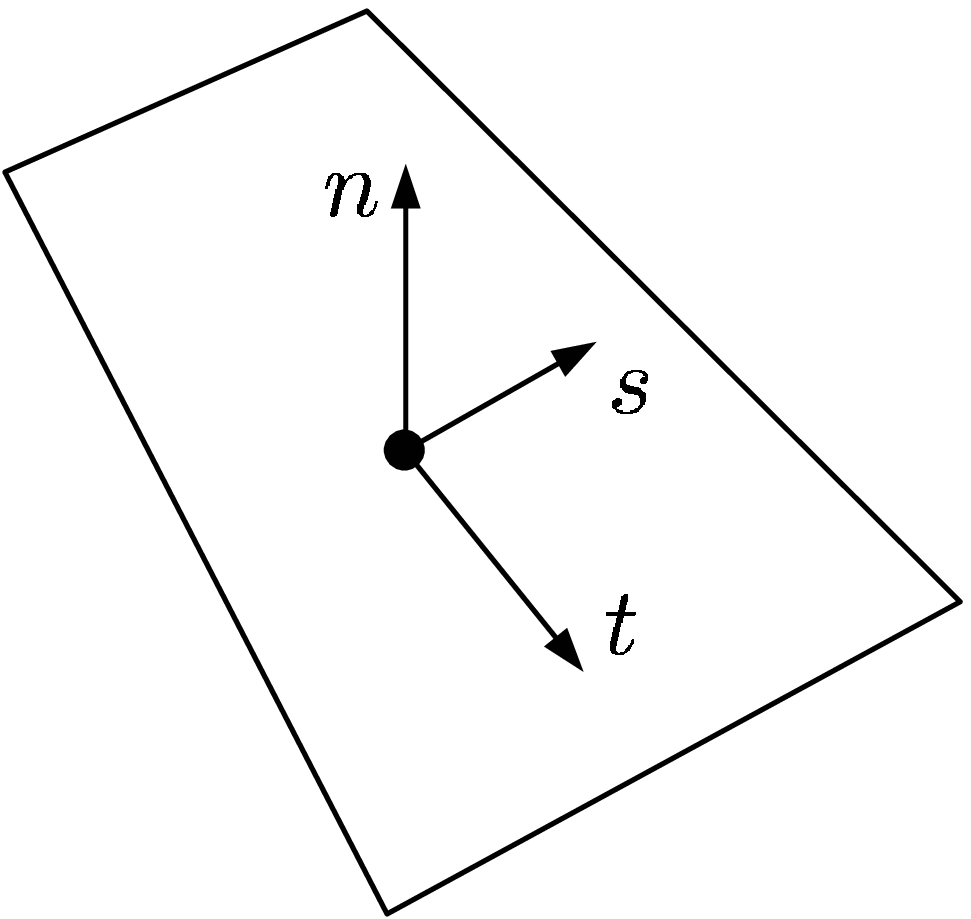, width=5cm}\\
(a)\\
\epsfig{file=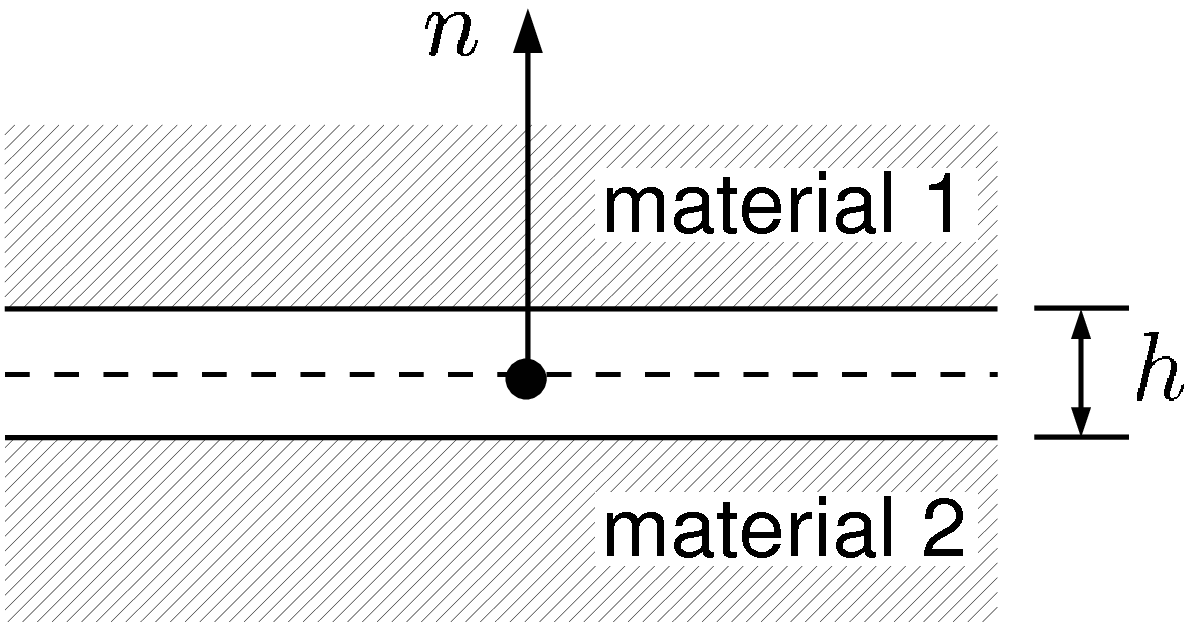, width=5cm}\\
(b)
\end{tabular}
\end{center}
\caption{Interface. (a) Tangential plane of interface with the local coordinate system $n$, $s$ and $t$. (b) Cross-section of interface of thickeness $h$.}
\label{fig:interface}
\end{figure}

In the lattice model, the nodal degrees of freedom are related to displacement jumps at the middle cross-section of the lattice element.
This three-dimensional displacement jump $\mathbf{u}_{\rm c} = \left(u_{\rm n}, u_{\rm s}, u_{\rm t}\right)^T$ is transformed into strains $\boldsymbol{\varepsilon} = \left(\varepsilon_{\rm n}, \varepsilon_{\rm s}, \varepsilon_{\rm t} \right)^T$ by means of the interface thickness $h$ as
\begin{equation}\label{eq:smear}
\boldsymbol{\varepsilon} = \dfrac{\mathbf{u}_{\rm c}}{h}
\end{equation}
The three subscripts $n$, $s$ and $t$ denote the normal and two tangential directions in the local coordinate system of the interface (Figure~\ref{fig:interface}a).
The thickness of the interface $h$ (Figure~\ref{fig:interface}b) is chosen to be equal to the lattice elements crossing the interface between reinforcement steel and concrete.
The strains are related to the nominal stress $\boldsymbol{\sigma} = \left(\sigma_{\rm n}, \sigma_{\rm s}, \sigma_{\rm t} \right)^T$ by the elasto-plastic stress-strain relationship
\begin{equation}\label{eq:stressStrain}
\boldsymbol{\sigma} = \mathbf{D}_{\rm e} \left(\boldsymbol{\varepsilon}- \boldsymbol{\varepsilon}_{\rm c} - \boldsymbol{\varepsilon}_{\rm p}\right)
\end{equation}
where $\mathbf{D}_{\rm e}$ is the elastic stiffness, $\boldsymbol{\varepsilon}_{\rm c} = \left(\varepsilon_{\rm c}, 0, 0 \right)^T$ is the eigenstrain describing the expansion of the corrosion product and $\boldsymbol{\varepsilon}_{\rm p} = \left(\varepsilon_{\rm pn}, \varepsilon_{\rm ps}, \varepsilon_{\rm pt} \right)^T$ is the plastic strain.

\begin{figure}
\begin{center}
\epsfig{file=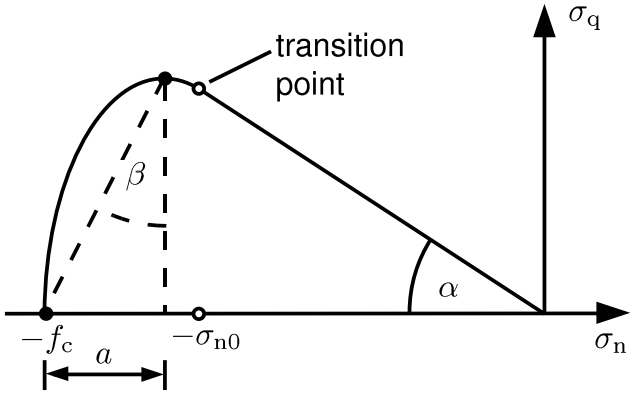,width=8.5cm}
\end{center}
\caption{Yield surface: Mohr-Coulomb friction law combined with a cap.}
\label{fig:surface}
\end{figure}

The yield surface of the plasticity model consists of a Mohr-Coulomb friction law combined with an elliptical cap. 
The shape of the cap surface is adjusted so that a smooth transition between the two surfaces is obtained (Figure~\ref{fig:surface}). This combination was initially proposed by \shortciteN{SwaSeo00} and further developed by \shortciteN{DolIbr07}.
The yield function $f$ depends on the normal stress $\sigma_{\rm n}$ and the shear stress norm $\sigma_{\rm q} = \sqrt{\sigma_{\rm s}^2 + \sigma_{\rm t}^2}$ as
\begin{small}
\begin{equation}\label{eq:yield}
f = \left\{ \begin{array}{ll} 
  \sigma_{\rm q} + \alpha \sigma_{\rm n} &  \mbox{if $\sigma_{\rm n0} \leq \sigma_{\rm n}$} \vspace{0.5cm}\\
  \sigma_{\rm q}^2 + \dfrac{\left(\sigma_{\rm n} + f_{\rm c} -a \right)^2}{\beta^2} - \dfrac{a^2}{\beta^2} & \mbox{if $\sigma_{\rm n} \leq \sigma_{\rm n0}$} \vspace{0.5cm}
\end{array} \right.
\end{equation}
\end{small}
\noindent where $\alpha$ is the frictional angle and $f_{\rm c}$ is the compressive strength.
Furthermore,
\begin{equation} 
a = \dfrac{\beta \alpha f_{\rm c}}{\alpha \beta + \sqrt{1+\beta^2\alpha^2}}
\end{equation}
where $\beta$ is the ratio of the short and long radii of the cap ellipse (Figure~\ref{fig:surface}).
At the point where the two parts of the yield surface meet, the normal stress is  $\sigma_{\rm n0} = -\dfrac{a}{ \beta \alpha \sqrt{1+ \beta^2 \alpha^2 }}$.
The rate of the plastic strains in Equation~\ref{eq:stressStrain} is
\begin{equation}\label{eq:flow}
\dot{\boldsymbol{\varepsilon}}_{\rm p} = \dot{\lambda} \dfrac{\partial g} {\partial \bar{\boldsymbol{\sigma}}}
\end{equation}
where $g$ is the plastic potential and $\lambda$ is the plastic multiplier.
In the present study, $g$ is chosen to be very similar to the yield function $f$. 
The only difference is that $\alpha$ is replaced by the dilatancy angle $\psi$ so that the magnitude of the normal plasticity strain generated during shear loading can be controlled.
The plasticity model is completed by the loading and unloading conditions:
\begin{equation}\label{eq:loadUn}
f \leq 0 \mbox{,} \hspace{0.5cm} \dot{\lambda} \geq 0 \mbox{,} \hspace{0.5cm} \dot{\lambda} f = 0
\end{equation}
This plasticity bond model is similar to the one developed by \shortciteN{Lun05a}.
However, in the present work, the response is assumed to be perfectly plastic, i.e. the shape of the yield surface is independent of the plastic strains. The bond model will be extended to hardening and softening in future work. The implementation of the present model is simplified by introducing a smooth transition between the cap and the frictional law. Thus, a special vertex stress return in the transition region is not required.

\subsection{\em Model for corrosion between concrete and reinforcement}
\begin{figure}
\begin{center}
\begin{tabular}{c}
\epsfig{file=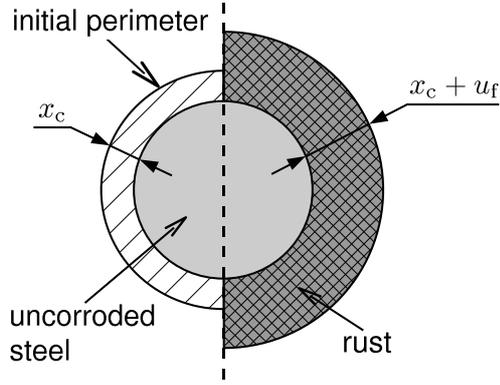,width=6.5cm}
\end{tabular}
\end{center}
\caption{Representation of the corrosion process as an expansive layer of rust.}
\label{fig:corrosion}
\end{figure}

The effect of corrosion is idealised as an eigenstrain $\varepsilon_{\rm c}$, which is determined from the free expansion of the corrosion product $u_{\rm c}$ as $\varepsilon_{\rm c} = u_{\rm c}/h$ (Figure~\ref{fig:corrosion}). 
This expansion is related to the corrosion penetration depth $x_{\rm c}$ as
\begin{equation}\label{eq:corExp}
u_{\rm c} = \sqrt{r^2 + \left(2 r x_{\rm c} - x_{\rm c}^2\right)\left(\lambda_{\rm c} - 1\right)} - r
\end{equation} 
where $\lambda_{\rm c}$ is the ratio of the volume ratio of rust and steel. The corrosion penetration $x_{\rm c}$ is related to the corrosion percentage $\rho_{\rm c}$ as
\begin{equation}\label{eq:corPen}
x_{\rm c} = r \left(1-\sqrt{1-\dfrac{\rho_{\rm c}}{100}}\right) 
\end{equation}

\section{COMPARISON WITH EXPERIMENTAL RESULTS}

\begin{figure}[ht!]
\begin{center}
\begin{tabular}{c}
\epsfig{file=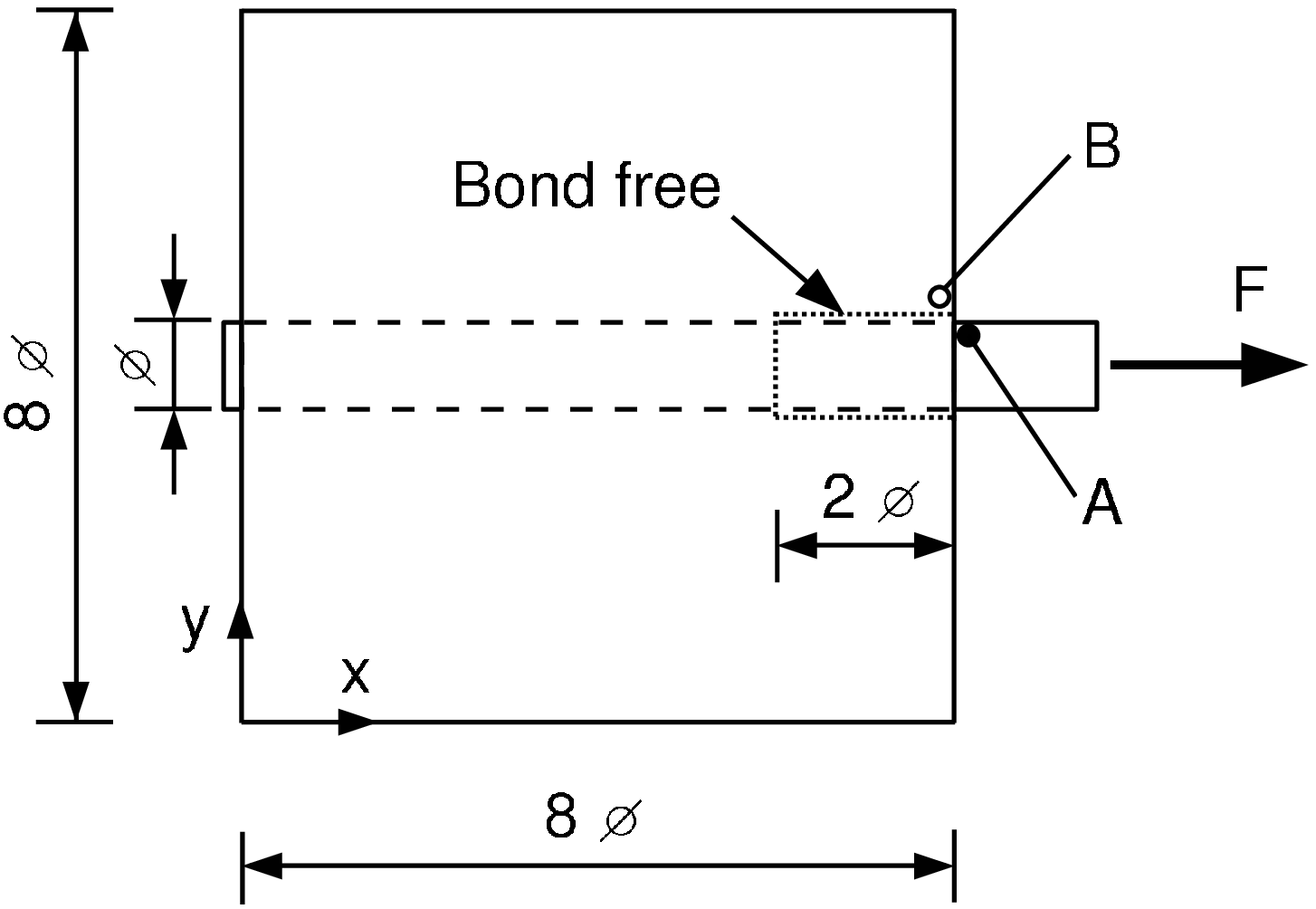,width=7.5cm}\\ 
(a)\\
\epsfig{file=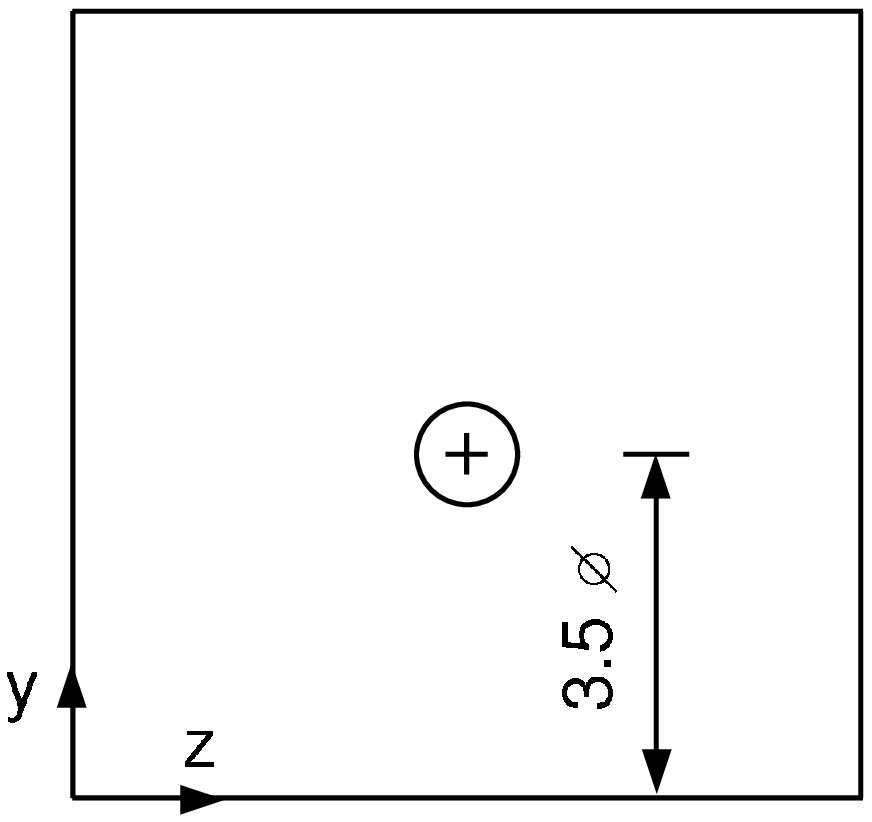,width=5.5cm}\\
(b)
\end{tabular}
\end{center}
\caption{Geometry and loading set-up for the by corrosion pull-out test reported by \protect \shortciteN{LeeNogTom02}. The reinforcement bar with a diameter $\diameter = 13$~mm  is placed eccentrically in $y$-direction in the concrete specimen. No lateral reinforcement is provided.}
\label{fig:geometry}
\end{figure}

The lattice approach is used to model the experiments reported by \shortciteN{LeeNogTom02}.
The geometry and loading setup of the experiments is shown in Figure~\ref{fig:geometry}.
Reinforcement bars ($\diameter = 13$~mm) embedded in concrete cubes were initially subjected to corrosion and subsequently pulled out.

\begin{figure}
\begin{center}
\epsfig{file=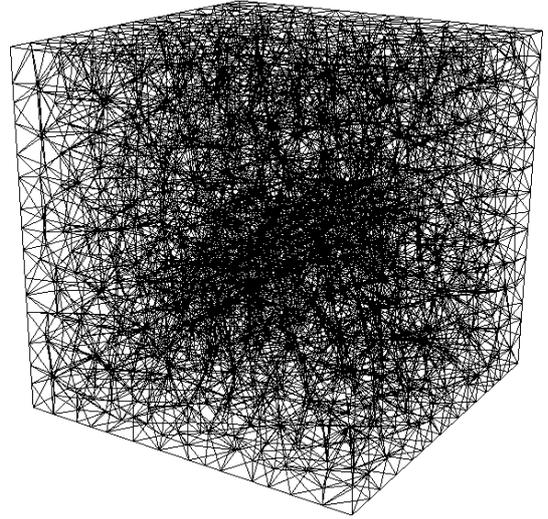,width=7cm}
\end{center}
\caption{Mesh for the lattice analysis.}
\label{fig:mesh}
\end{figure}

The concrete used in the experiments is characterised by a Young's modulus of $E_{\rm c} =22.6$~GPa, Poisson's ratio of $\nu_{\rm c}=0.17$, a tensile strength of $f_{\rm t} = 2.7$~MPa and a compressive strength of $f_{\rm c} = 24.7$~MPa. The Young's modulus of the reinforcement is $E_{\rm s} = 183$~GPa.
In the present study ,the response of concrete, reinforcement and bond between concrete and reinforcement is modelled by the lattice approach described earlier. 
The lattice for the analyses is shown in Figure~\ref{fig:mesh}. For the reinforcement and the interface between reinforcement and corrosion, the mesh is structured. For the concrete the lattice is random.

\begin{figure}
\begin{center}
\epsfig{file=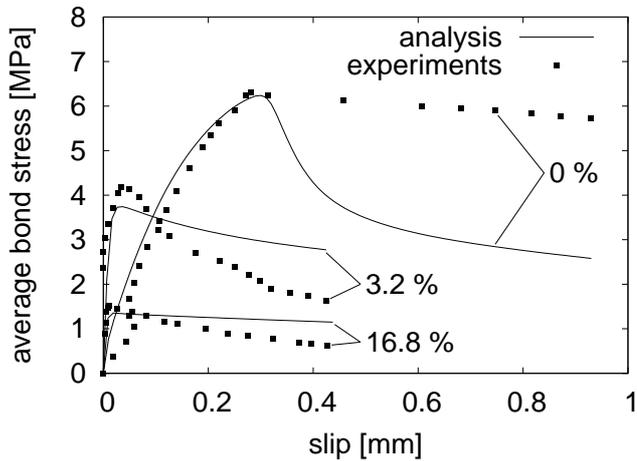,width=9cm}
\end{center}
\caption{Comparison of predicted average bond stress-slip curves and experimental data reported by \protect \shortciteN{LeeNogTom02} for three corrosion percentages $\rho_{\rm c} = 0$, $3.2$ and $16.8$~\%.}
\label{fig:ld}
\end{figure}

Three analyses were performed.
In the first analysis the reinforcement was pulled out without initial corrosion.
In the other two analyses, corrosion percentages of $\rho_{\rm c} = 3.2$ and $16.8$~\% were considered before the pullout. 
Assuming uniform corrosion, the corrosion percentages were transformed according to Equation~\ref{eq:corPen} to corrosion penetrations of $x_{\rm c} = 105$ and $571$~$\mu$m. With an expansion ratio of $\lambda_{\rm c} = 1.4$, this gives, according to Equation~\ref{eq:corExp}, free corrosion product expansions of $u_{\rm c} = 41.6$ and $214.8$~$\mu$m, respectively.
With an interface element thickness of 1 mm, this gives eigenstrains of $\varepsilon_{\rm c} = 0.0416$ and $0.2148$.
In all three analyses, the load $F$ was controlled by the end slip in the form of relative horizontal displacements of nodes $A$ and $B$ as shown in Figure~\ref{fig:geometry}a.
The results of the analyses are compared to the experimental results in the form of average bond stress-slip curves shown in Figure~\ref{fig:ld}. Here, the average bond stress was determined as $\tau = F/(\pi \diameter \ell)$, where $\ell = 6 \diameter$ is the embedded length.

The pre-peak regime of the load-split curves obtained in the analyses is in very good agreement with the experiments. 
However, the post-peak responses obtained in the analyses deviate considerably from those reported in the literature. 
In particular, the analysis of the corrosion free case exhibits a more brittle response than reported in the experiments.
On the other hand, the load-slip curves with initial corrosion exhibit a more ductile response than reported in the experiments. 
This is surprising, since it is expected that cracking in the concrete cover should reduce the pressure at the interface and, thus, also the tangential stresses. Consequently, a more consistent pattern of post-peak response might be expected across all three cases.
More studies are required to explore this observation.

\begin{figure}[ht]
\begin{center}
\begin{tabular}{c}
\epsfig{file=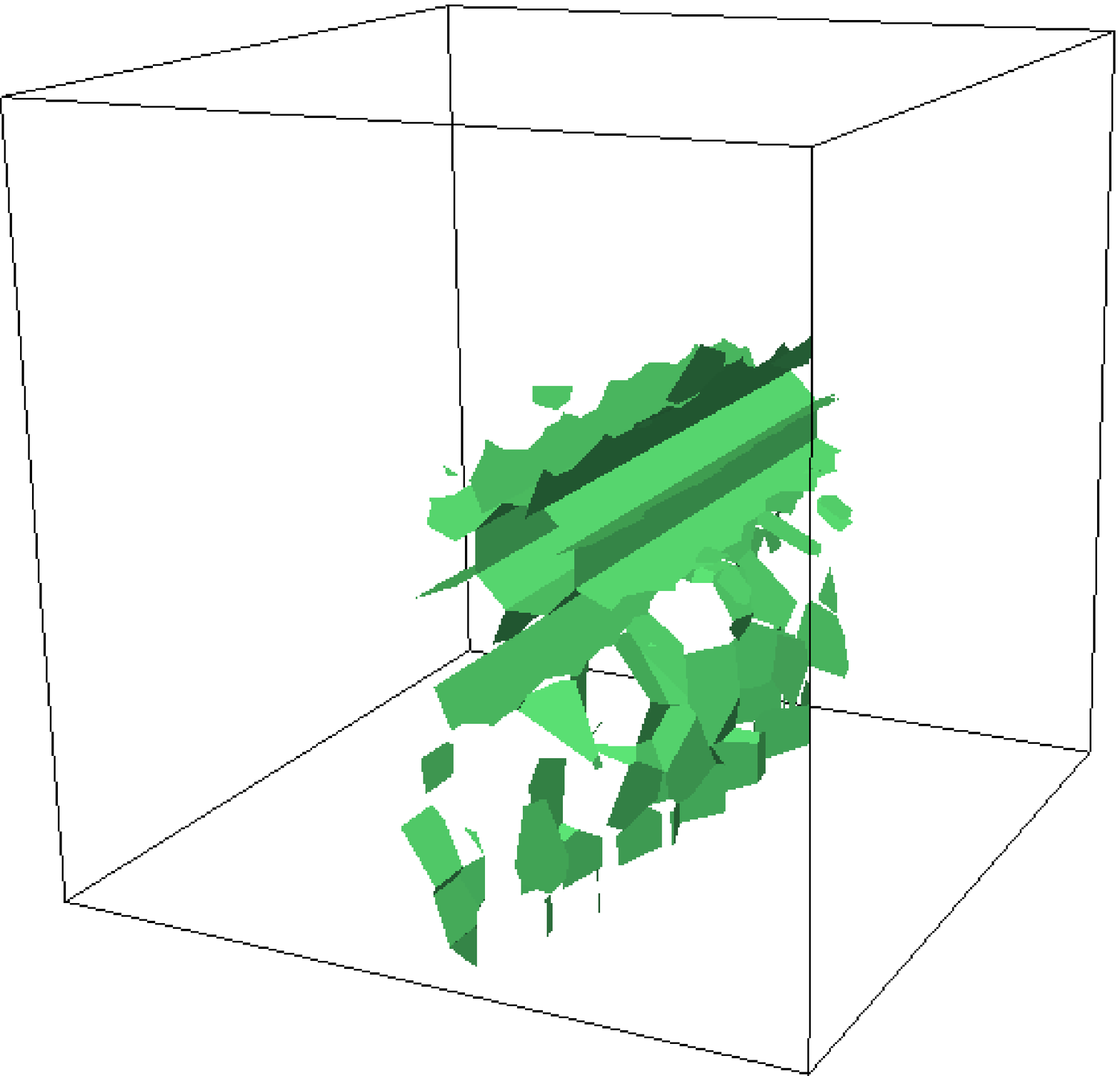,width=6cm}\\ 
(a)\\
 \epsfig{file=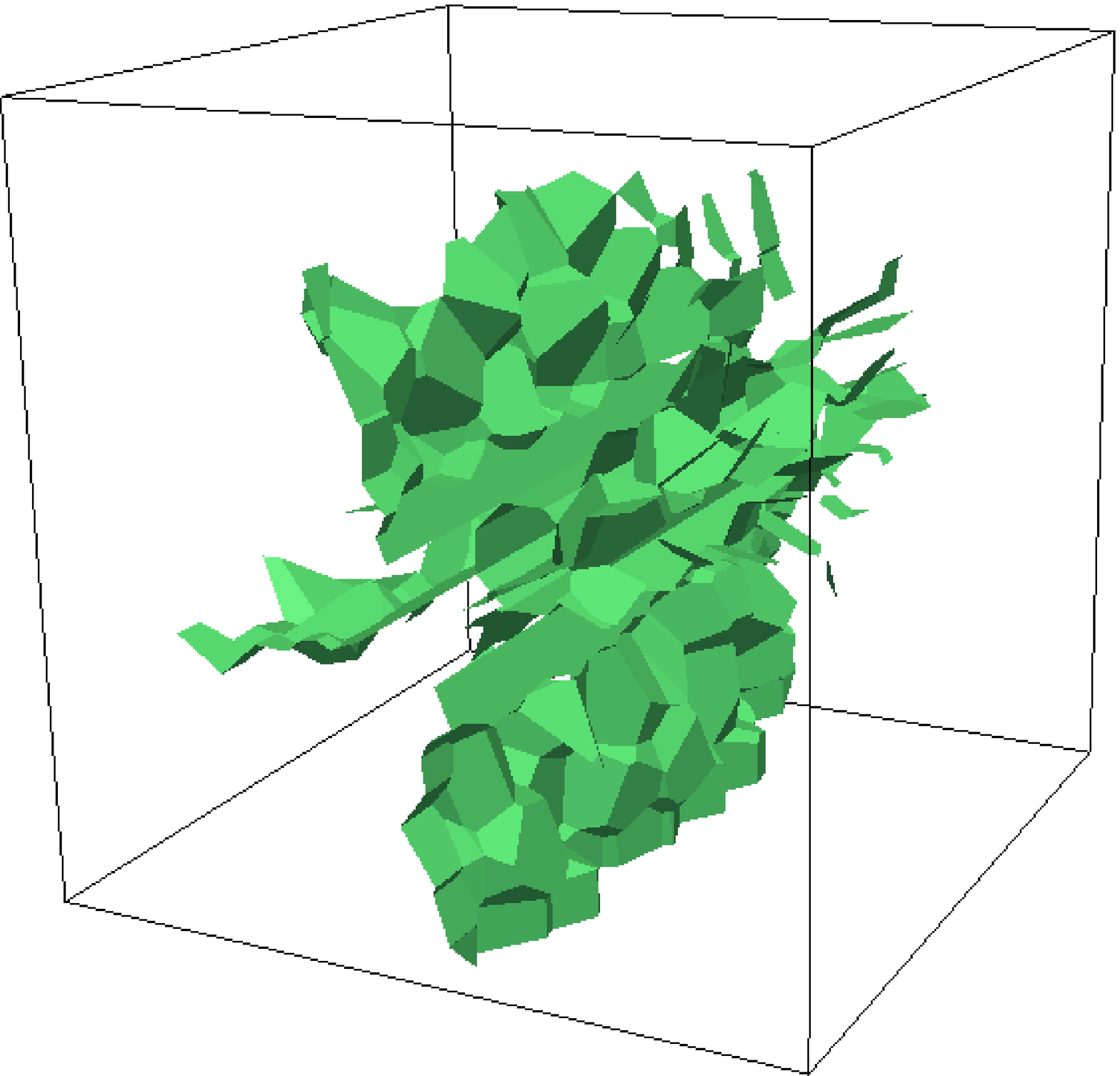,width=6cm}\\
(b)\\
\end{tabular}
\end{center}
\caption{Crack patterns for the pullout analysis for the corrosion-free case at (a) peak and (b) end of the average bond stress slip curve. Cracks initiate at the interface between reinforcement and concrete and propgate to the specimen surface.}
\label{fig:crack1}
\end{figure}

For the analyses without corrosion, the crack patterns for the peak bond stress and the maximum slip (presented in Figure~\ref{fig:ld}) are shown in Figure~\ref{fig:crack1}. Crack patterns are visualised as those middle cross-sections of lattice elements in which the norm of the crack opening vector is greater than $10$~$\mu$m and increasing.
Thus, only active cracks are presented. 

At the peak of the average bond stress-slip curve, the concrete cover is cracked at its thinnest section (Figure~\ref{fig:crack1}a).
With further slip, additional cracks initiate from the reinforcement and propagate radially into the specimen as shown in Figure~\ref{fig:crack1}b.
\begin{figure}[ht]
\begin{center}
\begin{tabular}{c}
\epsfig{file=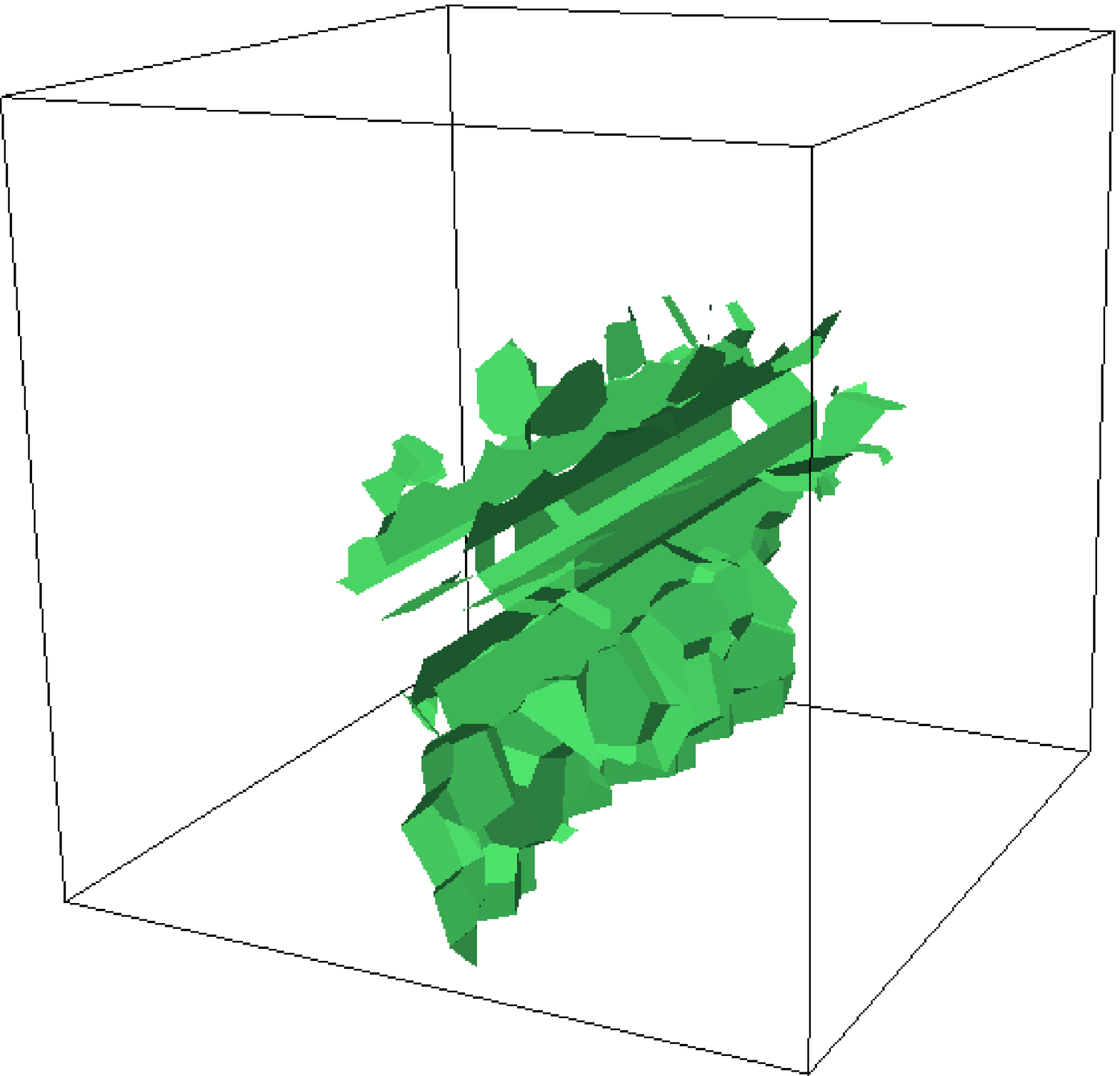,width=6cm}\\
(a) \\
\epsfig{file=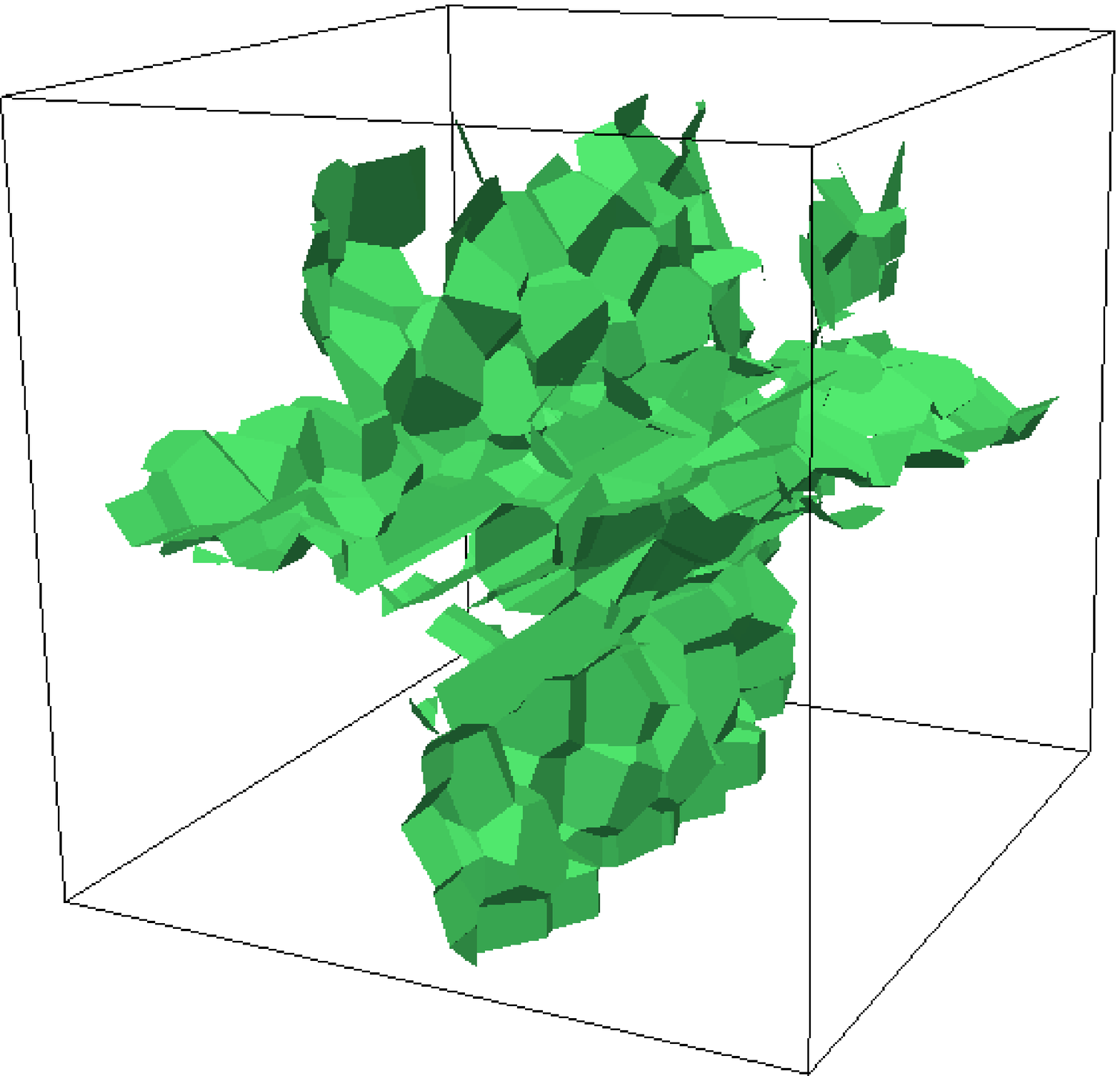,width=6cm}\\
(b)
\end{tabular}
\end{center}
\caption{Crack patterns for the analyses with (a) $3.2$~\% and (b) $16.8$~\% corrosion percentage before the pullout.}
\label{fig:crack2}
\end{figure}

In Figure~\ref{fig:crack2} the crack patterns are shown, for the two corrosion cases, at the end of the corrosion process.
For both corrosion cases, cracking of the concrete cover occurs before the pullout, which corresponds to the observations reported in the literature \shortcite{LeeNogTom02}. 

\section{CONCLUSIONS}

In the present work a lattice approach is used to describe the mechanical interaction of a corroding reinforcement
bar, the surrounding concrete and the interface between steel reinforcement and concrete. The cross-section of the ribbed reinforcement bar is taken to be circular, assuming that the interaction of the ribs of the deformed reinforcement bar and the surrounding concrete is included in a cap-plasticity interface model. 
This lattice approach is capable of representing many of the important characteristics of corrosion induced cracking and its influence on bond.
The idealisation of the corrosion expansion as an eigenstrain allows for the modelling of corrosion induced cracking. 
Furthermore, the frictional bond law can model the decrease of the bond strength if the concrete is pre-cracked.
Very good agreement with experimental results in the pre-peak regime of the bond stress-slip curves was obtained.
More studies are required to investigate the post-peak response of the bond stress-slip curves.
Also, further studies will be performed to investigate the influence of the element length of the interface between reinforcement and concrete on the analyses results. Also, we will study the influence of the stiffness of the lattice elements on corrosion induced cracking and its interplay with lateral confinement.

\section*{ACKNOWLEDGEMENTS}
The simulations were performed with the object-oriented finite element package OOFEM \shortcite{Pat99,PatBit01}, extended by the present authors.

\bibliographystyle{chicago}      % <---------------------------------- MOD
\begin{small}
\bibliography{general} % These is my BiBTeX reference list.
\end{small}
\end{document}